# TECHNOLOGY UPDATE OF A CONTROL SYSTEM TROUGH CORBA

A. Bertocchi, V. Bozzolan, G. Buceti, C. Centioli, F. Iannone, G. Mazza, M. Panella, V. Vitale, Associazione EURATOM-ENEA sulla Fusione, Centro Ricerche Frascati, Italy
D. Di Muzio, Università degli Studi "La Sapienza", Rome, Italy

## Abstract

After more than 10 years of activity, FTU (Frascati Tokamak Upgrade) has to face the problem of taking advantage of the new hardware and software technology saving all the investments yet done. So, for example, the 20 years old Westinghouse PLCs (communicating only through a serial line) have to be left in operation while a web-based tool has to be released for plant monitoring purposes. CORBA bus has shown to be the answer to set up the communication between old and new hardware and software blocks. On FTU the control system is based on Basestar, a Digital/Compaq software suite. We will describe how the real time database of Basestar is now accessed by Java graphic tools, LabView and, in principle, any package accepting external routines. An account will be also done in the data acquisition area where a Camac serial highway driver has been integrated in the new architecture still through the CORBA bus.

## 1 INTRODUCTION

From the first release in 1990, the control and data acquisition system is now a reliable subsystem of the whole FTU plant [1]. Even if we consider the control system as a black box, there are two 'old fashion' functionalities cannot be left unchanged: the user interface and the Camac access.

### 1.1 User interface

In the last years powerful tools have been developed and the old man machine interface is barely accepted from the users. The question is how to replace this last layer of the control system without affecting the functionality of the system core.

### 1.2 Camac access

Here the question is how to get rid of VAX machine preserving CAMAC access and the bulk of the data acquisition processes. Presently we are using a Kinetic serial highway driver (SHD) on the VAX/VMS Q-bus. We have been long time evaluating how to move from VAX to other systems, but the CAMAC link to Q-bus was a major constraint.

## 2 CONTROL SYSTEM ON CORBA

Our control system [2] is based on Compaq Basestar, a family of products that lay the foundations for a complete manufacturing integration solution: core platform of services, device connect capabilities, and a robust user environment comprised of graphics enabling and application programming tools. Based on the client-server architecture, the FTU control system was not designed to have location-independent communication services.

On the other end, middleware as CORBA is the technology that facilitates integration of components in a distributed system. CORBA allows elements of applications to interoperate across network links, despite differences in underlying communication protocols, system architectures, operating systems, databases, and other application services.

To make possible the use of tools like LabView or any Java user developed code on a PC over the net it needs just to make available on the net the control system API. This is just the job of CORBA services. Fig. 1 shows the changes in the upper layer of the control system. The CORBA server running on the node physically connected to the PLC serial lines provides the access to the PLC memory buffer. Under this architecture a survey of the plant can be done everywhere from the experts.

## 3 NET ACCESS TO CAMAC

As we mentioned before, the biggest issue is how to get rid of VAX machine preserving CAMAC access. At present we are using a Kinetic serial highway driver (SHD) on the VAX/VMS Q-bus. There were several options for a new SHD:
1. A commercial VME module; the driver was available only for VxWorks systems, while the VME standards used in our laboratories are LynxOS and Linux.
2. In house developed VME module. JET had developed its own VME SHD, but the support was not guaranteed at a commercial level.
3. SCSI SHD. This is quite a widespread solution, adopted in several laboratories; we found it expensive and slow.
4. PCI SHD.

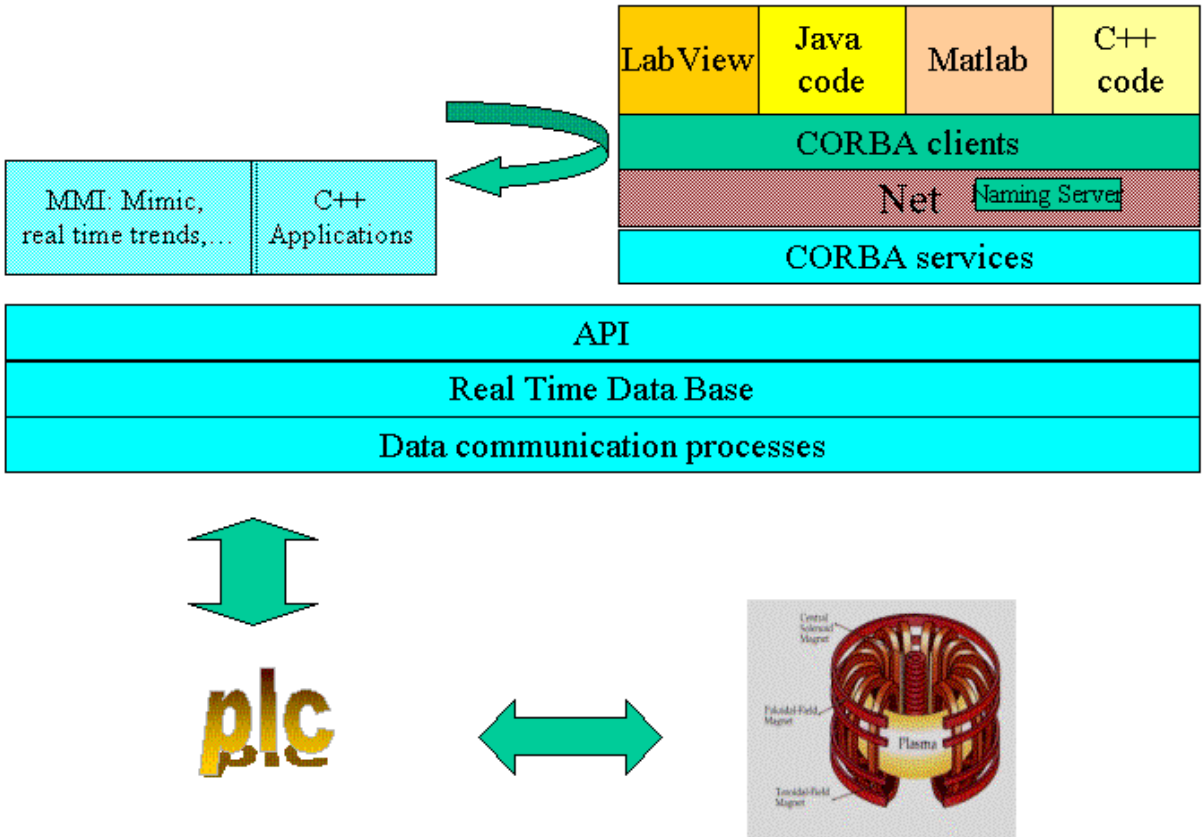

Figure 1: Sketch of the swap of the upper layer of the control system

The PCI SHD seemed to be the most up-to-date and economic solution. There were several modules available on the market, and we took into consideration the Kinetic System PCI SHD 2115, with support for Alpha/VMS, and the Hytec PCI SHD 5992 running a Windows/NT software driver. At first glance, the first solution seemed to be the most appealing: practically we could have ported CAMAC management directly on the Alpha machine, where the data acquisition system would have been running. On the other hand, due to our previous experience, we didn't want to limit long-term evolutions of the acquisition system with hardware control issues. So, we have moved to a modular architecture in which CAMAC loops will have been handled independently, like any other acquisition unit. The first step would be disentangle data acquisition processes and CAMAC control, running respectively on Alpha/VMS and a PC, equipped with the Hytec PCI SHD, using CORBA as a unifying layer (see fig. 2).

To prototype the new architecture, we realized a simple CORBA client-server structure distributed on three nodes, namely: a DEC Alpha/VMS, a Pentium III PC at 550 MHz, (operating system: Windows/NT 4.0), and a EUROCOM 138 VME CPU board, with an Intel Celeron CPU at 300 MHz (operating system: Linux Slackware 7.1) we exploited as CORBA name server. We installed OmniORB 3.0 as a CORBA server on the PC we had equipped with the Hytec SHD and defined the essential ESONE CAMAC access routines as CORBA services, in C++. A CORBA client has been defined on the Alpha machine on which data acquisition processes run. On the Alpha node the ESONE calls to CAMAC have been wrapped, the inner layer being an invocation of the relevant CORBA service; so that we replaced one library instead of the whole set of processes handling hardware modules. We have successfully tested CAMAC access via a general purpose crate initialization and dataway reading/writing routine. The time behavior of a net based CAMAC connection has also been tested; as expected, fastest response is achieved when the PC hosting the SHD and the clients are in the same subLAN, since any intervention of the router may heavily influence the data transfer throughput.

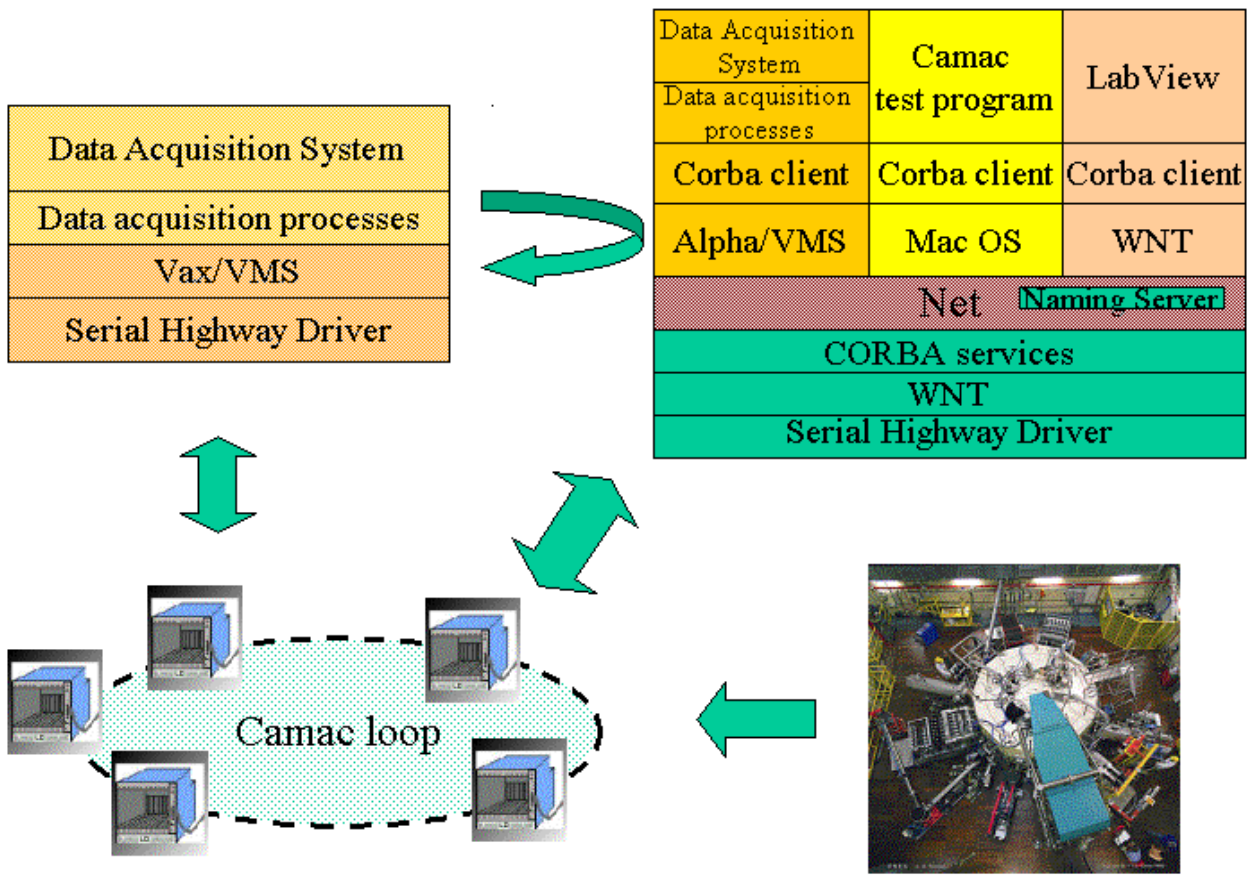

Figure 2: Migration to a Camac net access

In tab. 1 are listed the times for writing and reading 1 ksamples from a Camac dual port memory module in a loop of 40 crates. The results differ because the following setup:
1. Direct acquisition via Qbus on Vax
2. Acquisition trough CORBA server on LAN
3. Acquisition trough CORBA server with Ethernet direct connection (inverted cable)

The Ethernet connection is at 100 Mb/s. The result shows that, because the large Ethernet bandwidth and the faster processor (compared with the Vax), there is no significant increase in the acquisition time.

Table 1: Camac access time for 1 k samples (ms)

| Configuration | Write | Read |
|---|---|---|
| Qbus | 440 | 410 |
| LAN connection | 520 | 512 |
| Direct connection | 508 | 500 |

## 4 CONCLUSIONS

Distributed object technology has deeply changed the way to manage legacy systems. In the area of the control systems, where a lot of hardware and related drivers has to be maintained for long time, the opportunity to wrap the lowest layer can significantly change the life of the whole system. In particular, on FTU, both the slow data acquisition via PLC and the fast data acquisition via Camac have been wrapped to allow data access with the most updated available tools.